\documentclass{article}
\usepackage{spconf,amsmath,graphicx,hyperref}
\usepackage{cite}
\usepackage{amsmath,amssymb,amsfonts}
\usepackage{algorithmic}
\usepackage{graphicx}
\usepackage{textcomp}
\usepackage{xcolor}
\usepackage{booktabs}
\usepackage{multirow}
\usepackage{subfloat}
\usepackage{subcaption}

\title{Comparative Study of Subjective Video Quality Assessment Test Methods in Crowdsourcing for Varied Use Cases}
\name{Babak Naderi, Ross Cutler}
\address{Microsoft Corporation, Redmond, USA}
\begin{document}
%
\maketitle
\begin{abstract}
In crowdsourced subjective video quality assessment, practitioners often face a choice between Absolute Category Rating (ACR), ACR with Hidden Reference (ACR-HR), and Comparison Category Rating (CCR). We conducted a P.910-compliant, side-by-side comparison across six studies using 15 talking-head sources of good and fair quality, processed with realistic degradations (blur, scaling, compression, freezing, and their combinations), as well as a practical bitrate-ladder task at 720p and 1080p resolutions. We evaluated statistical efficiency (standard deviations), economic efficiency, and decision agreement. Our results show that ACR-HR and ACR correlate strongly at the condition level, while CCR is more sensitive—capturing improvements beyond the reference. ACR-HR, however, exhibits compressed scale use, particularly for videos with fair source quality. ACR-HR is approximately twice as fast and cost-effective, with lower normalized variability, yet the choice of quality measurement method shifts saturation points and bitrate-ladder recommendations. Finally, we provide practical guidance on when to use each test method.
\end{abstract}

\begin{keywords}
Video quality assessment, subjective tests, ACR-HR, CCR, crowdsourcing
\end{keywords}
\section{Introduction}
\label{sec:intro}
Subjective video quality assessment (VQA) remains the gold standard for evaluating end-user experience in communication, streaming, and user-generated content pipelines. While laboratory studies provide strong internal validity, crowdsourcing systems have been shown to be accurate, reliable, faster, and cheaper \cite{hossfeld_best_2014, naderi_crowdsourcing_2024}. Yet, practitioners still face a persistent choice among test paradigms with different trade-offs: single-stimulus ACR, ACR-HR, and double-stimulus like DCR and CCR. These methods differ in stimulus presentation, presence of an explicit or implicit reference, rating scale, and per-subject time; consequently, they can yield different confidence intervals (CIs), costs, and recommendations—particularly when content quality varies widely. 

This paper presents a systematic, side-by-side study of ACR-HR and CCR in a crowdsourced setting using P.910-Crowd \cite{naderi_crowdsourcing_2024}. Using 15 talking-head source clips spanning good and fair qualities, we synthesize a broad set of realistic degradations (e.g., blur, scaling, compression, frame freezing and their combinations) and run six independent studies to quantify how test choice affects conclusions at both clip and condition levels. We further analyze a practical Real-time Communication (RTC) decision problem—constructing a bitrate ladder across 720p and 1080p resolutions—where small perceptual differences can materially change operating points. Our evaluation emphasizes statistical efficiency (standard deviation size), economic efficiency (participant time and monetary cost), and decision agreement across methods. 

Our contributions are: (1) a controlled, P.910-compliant crowdsourced comparison of ACR-HR and CCR over diverse, talking-head-centric impairments; (2) empirical guidance on when hidden references (ACR-HR) sufficient versus when  comparison rating (CCR) is warranted; and (3) an application study showing how method choice influences bitrate-ladder recommendations for RTC. Together, these results provide actionable guidance for researchers and practitioners choosing a subjective test design under budget, timeline, and sensitivity constraints.

\section{Related work}
Across domains, ACR (5-point) and ACR-HR are generally the most time- and cost-efficient general-purpose choices: large cross-method correlations, small confidence intervals (CIs) in many conditions, and the shortest per-stimulus durations—especially for IPTV/mobile HD and QVGA H.264 content~\cite{tominaga_performance_2010}; guidance surveys similarly emphasize ACR’s speed and precision~\cite{cermak_relationship_2011}. In broadcast/TV systems with MPEG impairments, a meta-experiment showed single-stimulus continuous tests can match CCR/Double Stimulus Continuous Quality Scale (DSCQS) accuracy after proper post-processing, i.e., no inherent accuracy edge for double-stimulus, only a sensitivity/throughput trade-off~\cite{pinson_comparing_2003}. For 2D and stereoscopic 3D 1080p H.264, DCR/DSIS achieved smaller CIs than ACR—especially at lower quality—while ACR remained the fastest; DSCQS was best only at very high quality~\cite{kawano_performance_2014}.  In VR with 3D graphics, DSIS/SAMVIQ outperformed ACR-HR on accuracy and DSIS achieved the highest accuracy in the shortest time, highlighting the value of an explicit reference when viewers lack strong priors~\cite{nehme_comparison_2020}.  In 360° VR video on HMDs, results are mixed: a direct comparison found M-ACR (two-pass ACR) slightly smaller CIs and less simulator sickness than DSIS, though the two methods were highly correlated and DSIS yielded slightly higher MOS at higher bitrates~\cite{singla_comparison_2018, singla_subjective_2019}.  A recent 360° dataset also validates that ACR and DCR are suitable baselines for HMD testing~\cite{elwardy_acr360_2023}.

\section{Methods}
\label{sec:method}

We evaluated the effect of typical video-related degradations on perceptual quality using different subjective test methods to identify discrepancies between the conclusions drawn from each test method. To this end, we created a dataset with typical artifacts introduced through video processing. We used two different quality levels for the source materials to observe interaction between that and the degradations. In addition, we focus on the construction of the bitrate ladder for real-time communication as a specific use case, showcasing the differences between the results of different subjective test methods. 

\subsection{Dataset}
The synthetic dataset was created using 15 talking-head videos from the VCD open-source dataset~\cite{naderi_vcd_2024} as source videos. These videos are user-generated content and were recorded with different devices, environments, participants, and lighting conditions. Of these, ten source videos belong to the good-quality group, with an average Mean Opinion Score (MOS) of 4.65, while five belong to the fair-quality group, with an average MOS of 3.4.The source videos were processed with the degradations listed in Table~\ref{table:deg}, each applied at multiple intensity levels. These degradations correspond to common restoration tasks, i.e. scaling (super-resolution), denoising, and deblurring, typical of RTC use cases.

Similarly, for the bitrate ladder use case, the same number of clips from each group was employed, with an average MOS of 4.1 for the good-quality group and 2.92 for the fair-quality group. In this use case, the analysis was restricted to 720p, 720p upscaled to 1080p, and 1080p resolutions with multiple bitrate categories per resolution resulting in an overall of 18 bitrate-resolution combinations. We have used H.264 for encoding the source videos with different target bitrates, and bilinear method for upscaling videos from 720p to 1080p.


\begin{table}[t]
\caption{Distortions used in Synthetic Dataset applied on source videos. Number of conditions shows number of intensity levels used.}
\label{table:deg} 
\centering
\resizebox{\columnwidth}{!}{
    \begin{tabular}{l c}
    \toprule
    \textbf{Distortions} & \textbf{N conditions}\\
    \midrule        
Blurring & 7 \\
Scaling (down and up) - Bilinear & 5 \\
Scaling down + DNN Super-resolution up-scaling & 7 \\
JPEG Compression & 10 \\
H.264 Quantization & 5 \\
Frame freezing & 9 \\
JPEG Compression $\times$ Scaling & 5$\times$5 \\
H.264 Quantization $\times$ Scaling & 3$\times$3 \\
Random Combinations $\times$ noise and color distortions & 10 \\
\midrule        
Overall & 87 \\

 \bottomrule    
\end{tabular}
}
\vspace{-0.3cm}
\end{table}

\subsection{Subjective Tests}
P.910-Crowd~\cite{naderi_crowdsourcing_2024} was employed to conduct multiple subjective video quality tests in accordance with ITU-T Rec. P.910\cite{itu-t_recommendation_p910_subjective_2023} via crowdsourcing. The test methods applied were Absolute Category Rating with Hidden Reference (ACR-HR) and Comparison Category Rating (CCR), which are briefly described below.

\textbf{ACR test method:} Participants rated video quality on a 5-point scale from Excellent (5) to Bad (1). Ratings were aggregated across clips and degradations to obtain MOS.

\textbf{ACR-HR test method:} Extends ACR by including reference videos in the test. Each session contained at least one reference within 13 rated clips in crowdsourcing test. Differential viewer scores (DV) were computed per subject and processed video sequence (PVS) using Equation~\ref{eq:dvs}, where V is the viewer’s ACR score~\cite{itu-t_recommendation_p910_subjective_2023}, and aggregated as DMOS.
\begin{equation}
\label{eq:dvs}
DV(pvs) = V(pvs) - V(ref) +5
\vspace{-0.1cm}
\end{equation}

\begin{table}[b]
\caption{Average number of reliable votes per condition with standard deviation and normalized deviation for the practical scale range. Datasets contain twice as many good-quality as fair-quality sources.}
\label{table:n_votes} 
\centering
\resizebox{\columnwidth}{!}{
    \begin{tabular}{l c c c c c c c }
    \toprule
    \multirow{2}{*}{\textbf{Dataset}} & \textbf{Reference} & \multicolumn{2}{c}{\textbf{ Avg. Votes}} & \multicolumn{2}{c}{\textbf{ Avg. Std.}} & \multicolumn{2}{c}{\textbf{ Avg. Nor. Std.}} \\
    & \textbf{quality}& \textbf{ACR-HR} & \textbf{CCR} & \textbf{ACR-HR} & \textbf{CCR} & \textbf{ACR-HR} & \textbf{CCR}\\
    \midrule        
Synthetic & Good & 83.4  & 79.9  & 0.78 & 0.69 & 0.19 &  0.23\\
 impairments & Fair & 40.0 & 41.2 & 0.73 & 0.81 & 0.18 &  0.27\\
 \midrule
\multirow{2}{*}{Bitrate ladder}  & Good & 206.5 & 181.5  & 0.84 & 0.90 & 0.21 &  0.30 \\
 & Fair & 88.4  &  89.2 & 0.81 & 0.87 & 0.20 &  0.29\\

 \bottomrule    
\end{tabular}
}
\end{table}

\begin{table*}[ht]
\caption{Correlation coefficients between opinion scores obtained from different subjective test methods using synthetic impairment dataset, aggregated in clip and degradation condition levels.}
\label{table:corr_syn} 
\centering
\resizebox{0.8\textwidth}{!}{
    \begin{tabular}{c l c c c c  c c c c}
    \toprule
        \multirow{2}{*}{\textbf{Level}} & \multirow{2}{*}{\textbf{Test method}} & \multicolumn{4}{c}{\textbf{Good Quality source}} &\multicolumn{4}{c}{\textbf{Fair Quality source}} \\
         & &  \textbf{Pearson} & \textbf{Spearman} &\textbf{Tau-b} & \textbf{Tau-b 95} & \textbf{Pearson} & \textbf{Spearman} &\textbf{Tau-b} & \textbf{Tau-b 95~\cite{majeedi_full_2023}} \\    
    \midrule        

 \multirow{4}{*}{\textbf{\rotatebox{90}{Clip}}} & ACR vs ACR-HR & 0.986 &	0.981	& 0.888	& 0.926	& 0.817	 & 0.789 &	0.609 &	0.662 \\
 & ACR vs CCR & 0.906	& 0.902	& 0.734&	0.773&	0.748&	0.684&	0.503&	0.554 \\
 & ACR-HR vs CCR & 0.907	& 0.906 &	0.736 &	0.773&	0.861&	0.804&	0.614&	0.68 \\
 & CCR r1 vs CCR r2 & 0.954	& 0.938 &	0.785&	0.829 &  \multicolumn{4}{c}{\textbf{--}} \\
 \midrule
\multirow{4}{*}{\textbf{\rotatebox{90}{Condition}}} & ACR vs ACR-HR & 1.0000 &	1.000 &	0.995 &	0.986 &	0.998 &	0.992 &	0.940 &	0.936 \\
 & ACR vs CCR & 0.970 & 0.974 &	0.866 &	0.901 &	0.961	 & 0.937&	0.795&	0.861 \\
 & ACR-HR vs CCR & 0.970 &	0.974 &	0.865 &	0.894 &	0.963 &	0.935 &	0.789 &	0.867 \\
 & CCR r1 vs CCR r2 & 0.992 &	0.990 &	0.920 &	0.941 &  \multicolumn{4}{c}{\textbf{--}} \\

 \bottomrule    
\end{tabular}
}
\end{table*}
\begin{figure*}[!ht]
    \centering
    \subfloat[]{\includegraphics[width=0.32\textwidth]{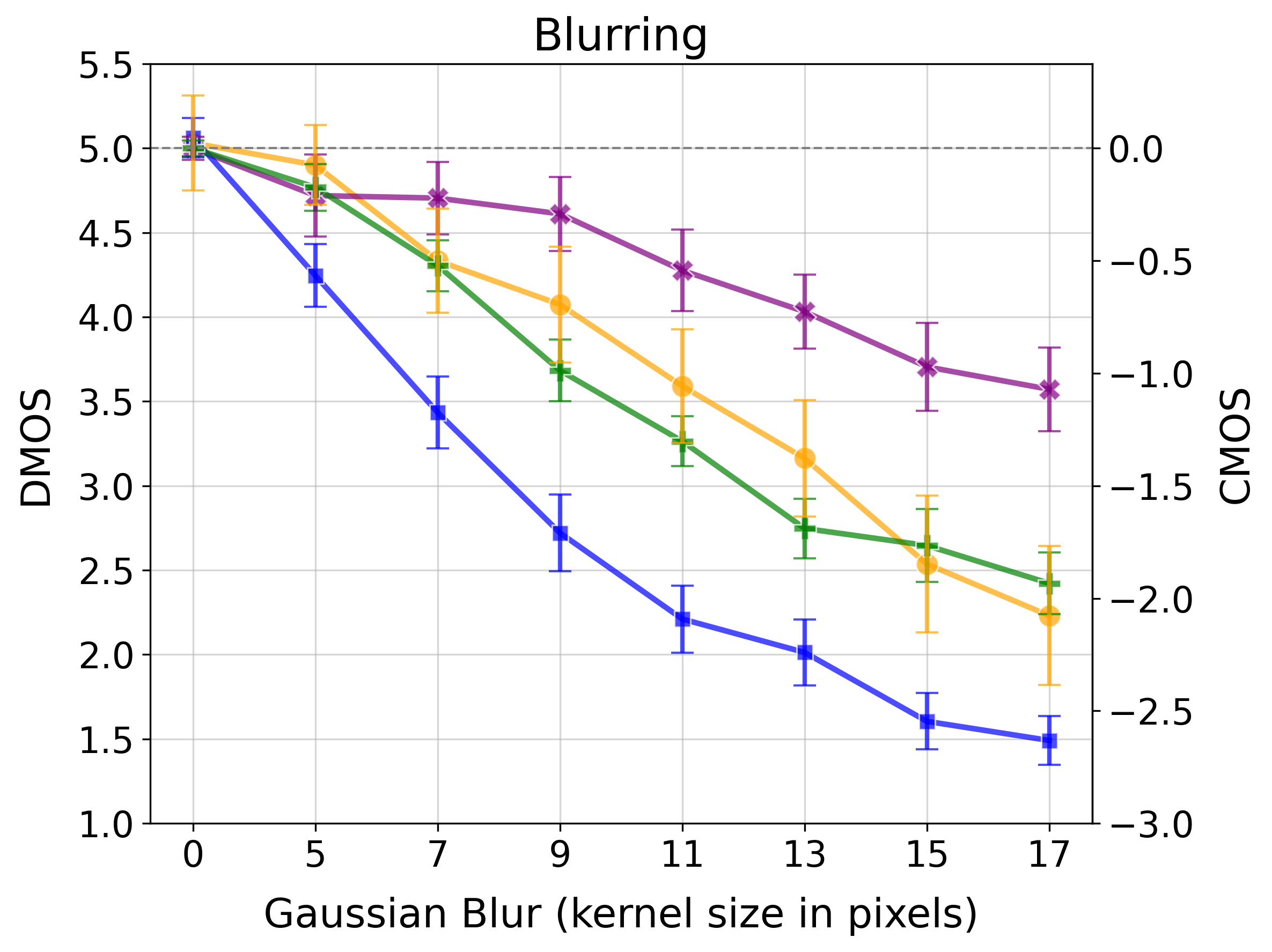}
    \label{fig:blur}}
    \hfill
    \subfloat[]{\includegraphics[width=0.32\textwidth]{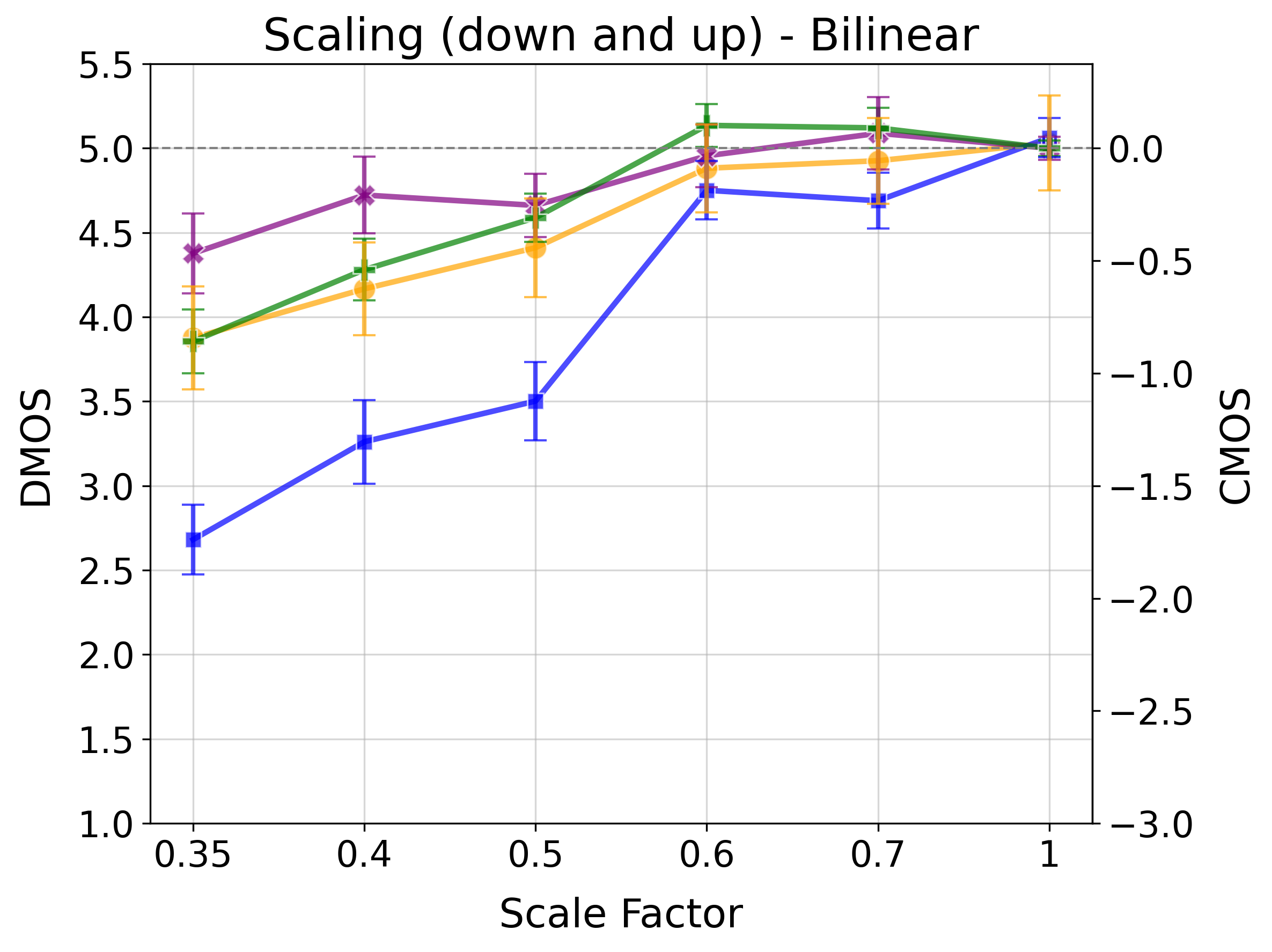}
    \label{fig:scale}}
    \hfill
    \subfloat[]{\includegraphics[width=0.32\textwidth]{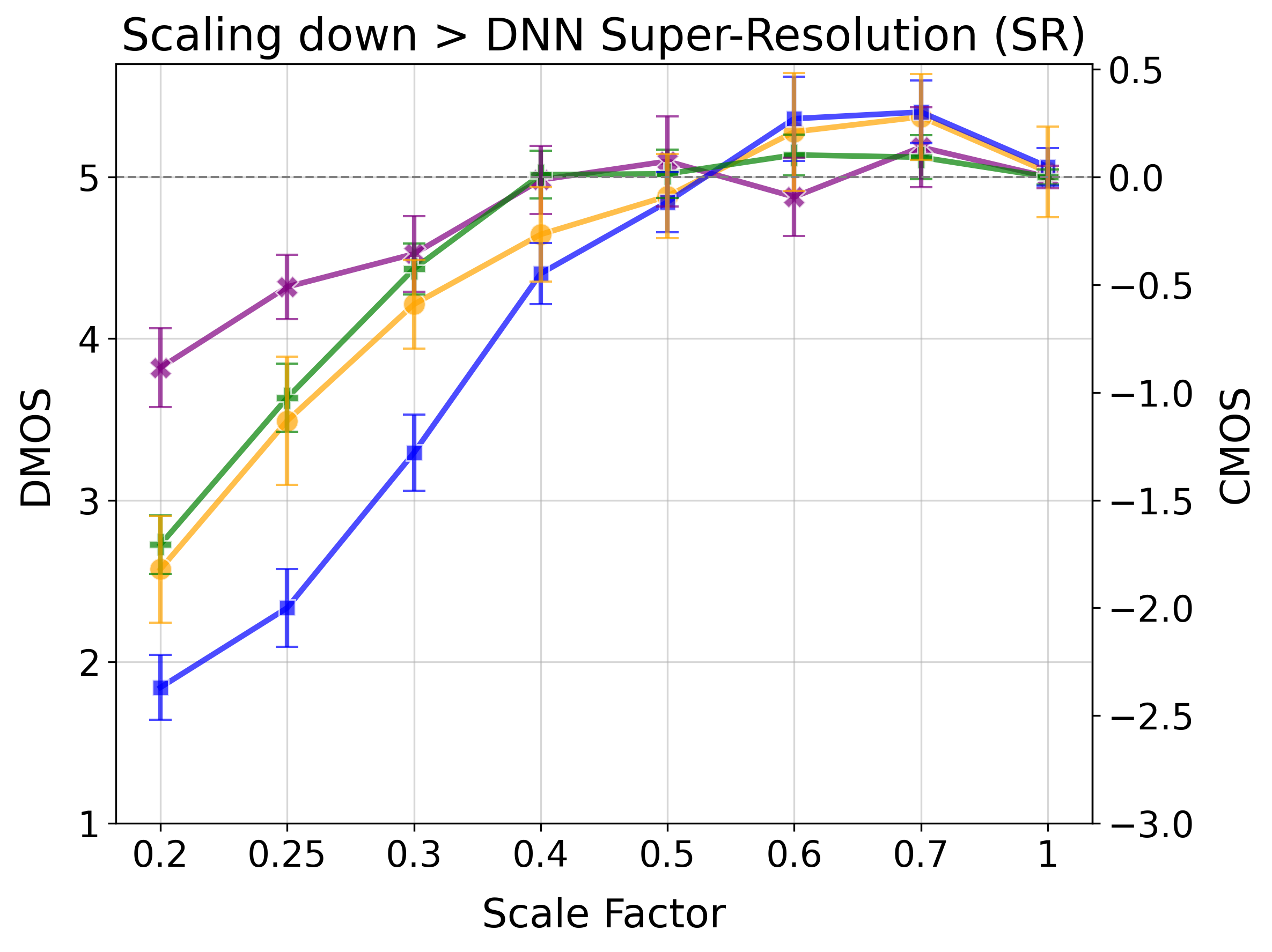}
    \label{fig:scale_dnn}}
    \hfill
    \subfloat[]{\includegraphics[width=0.32\textwidth]{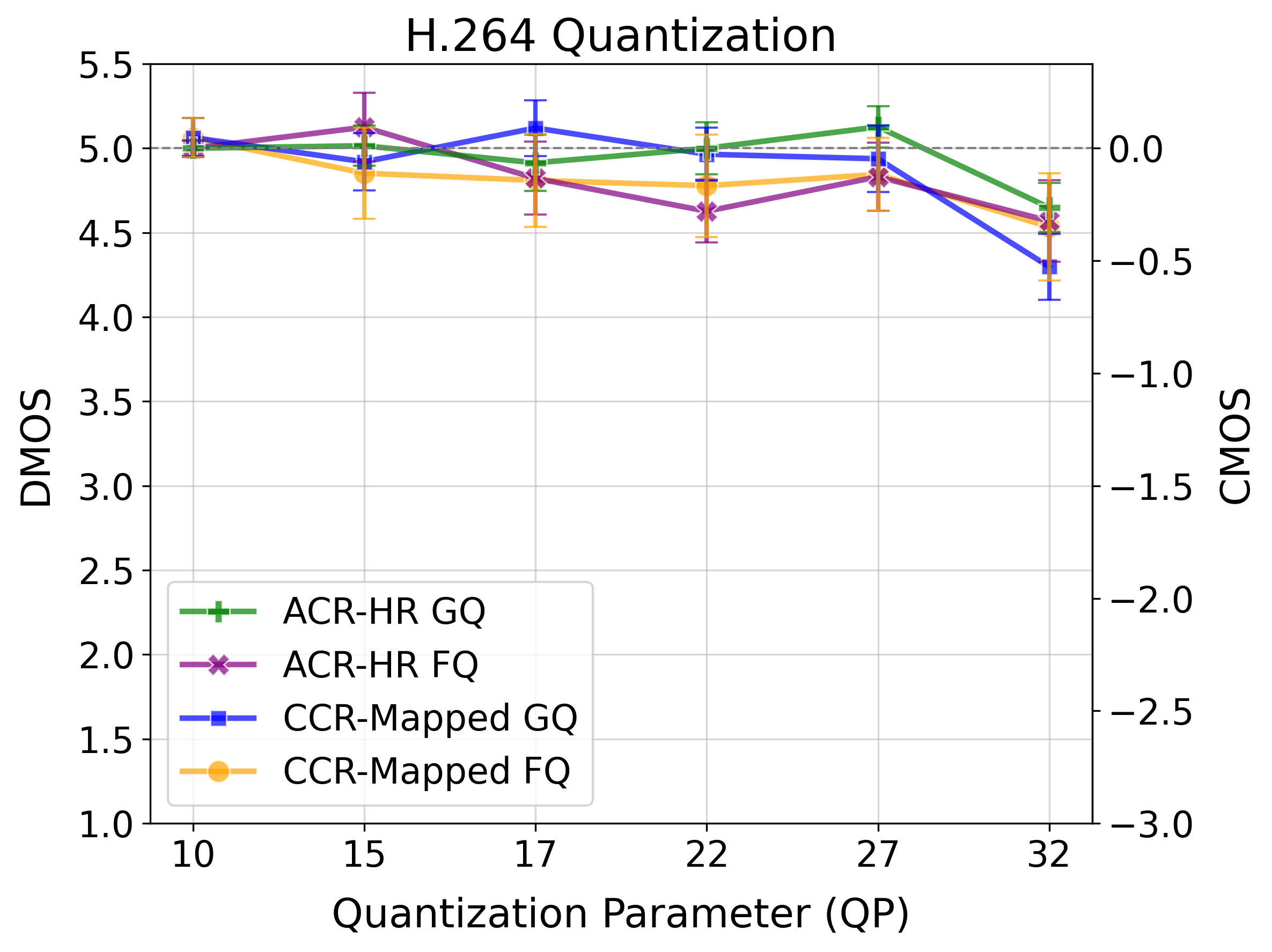}
    \label{fig:qp}}
    \hfill
    \subfloat[]{\includegraphics[width=0.32\textwidth]{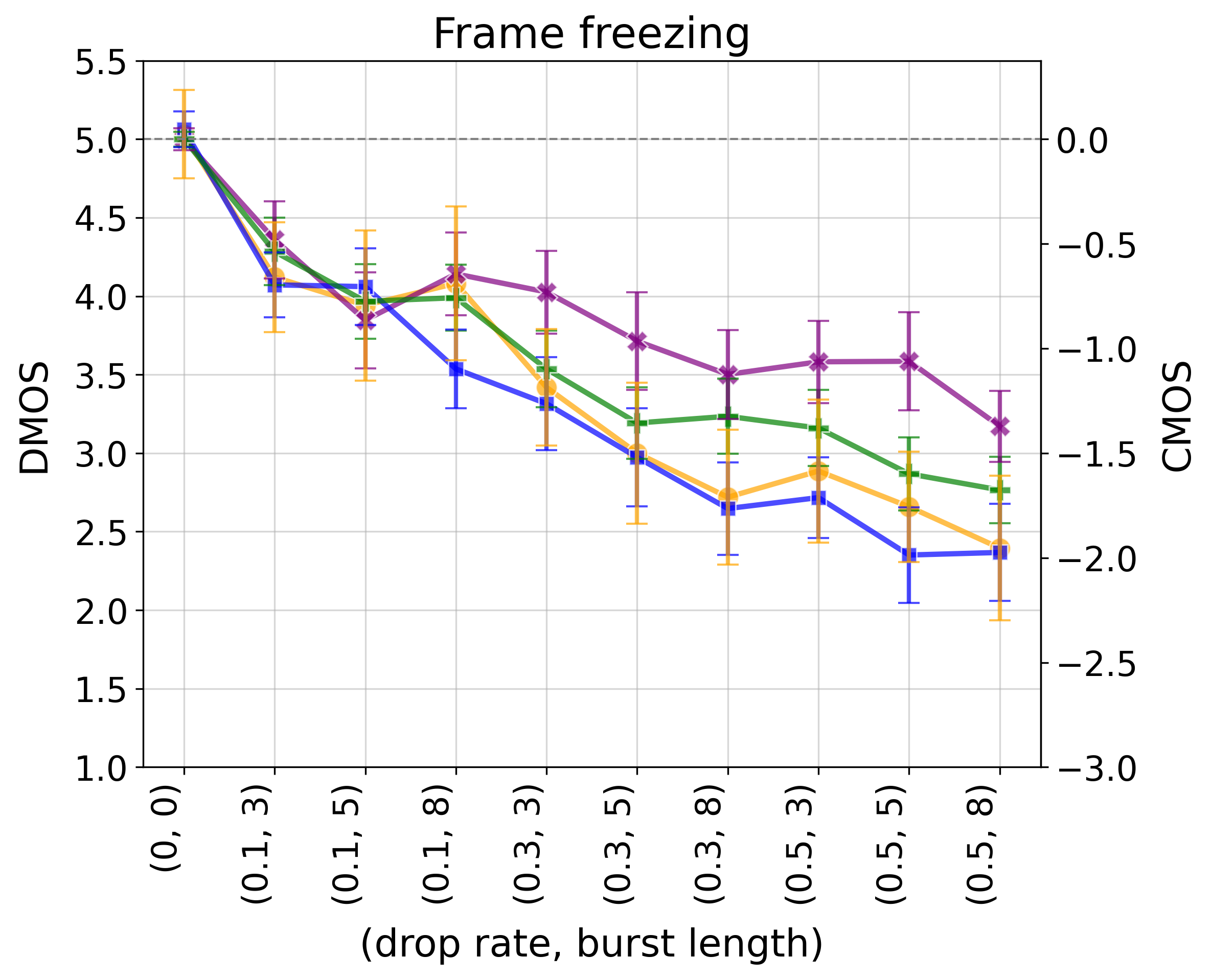}
    \label{fig:frame}}       
    \hfill    
    \subfloat[]{\includegraphics[width=0.32\textwidth]{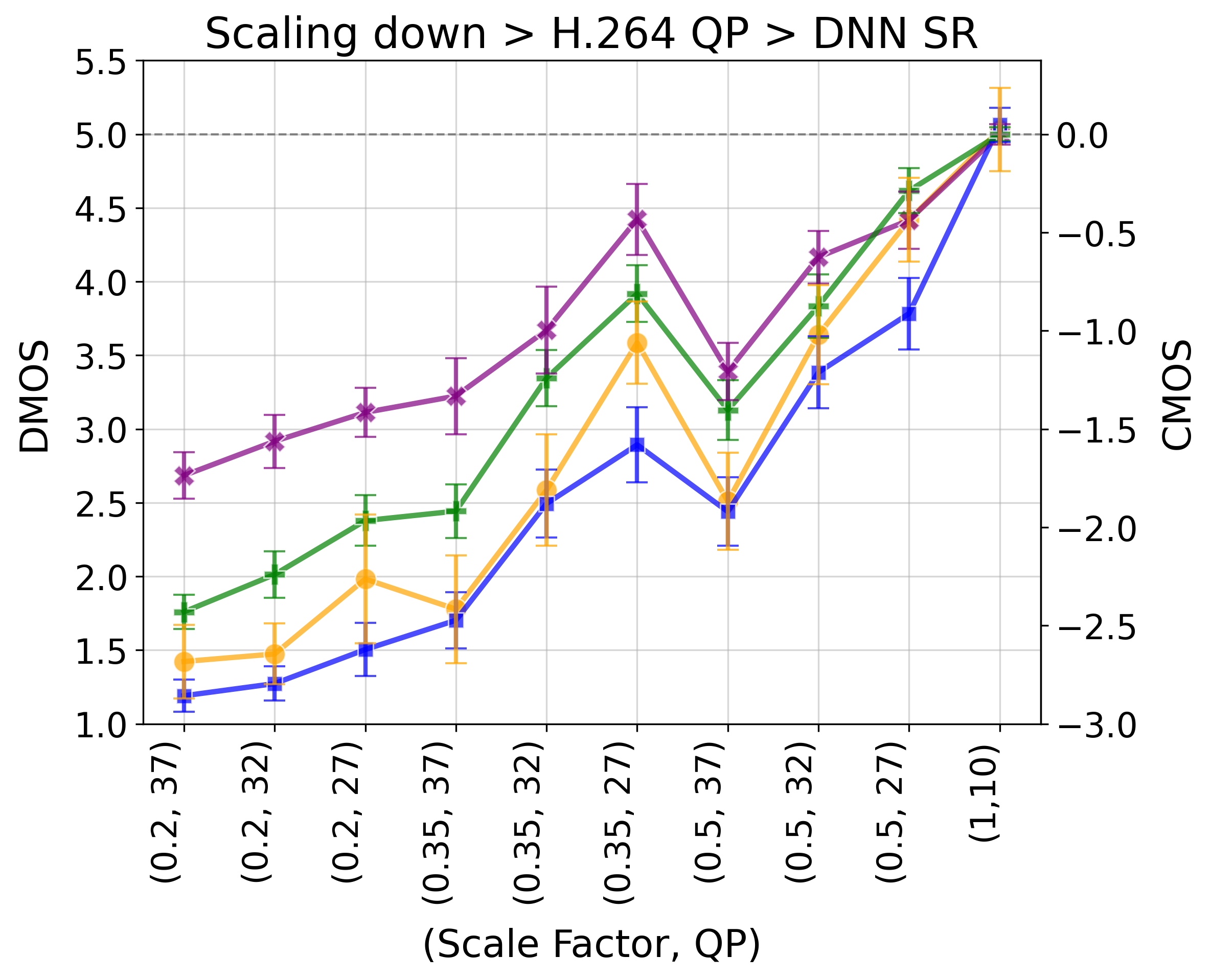}
    \label{fig:scale_dnn_qp}}

    \caption{Distortion–quality plots for selected degradations in the synthetic dataset. Quality was measured with ACR-HR and CCR using Good-Quality (GQ) and Fair-Quality (FQ) sources.}
    \label{fig:rd_plots}
\end{figure*}

\textbf{CCR test method:} Participants viewed the processed and reference videos in randomized order and rated the quality of the second video relative to the first on a 7-point scale, ranging from \textit{Much worse (-3)} through \textit{The same (0)} to \textit{Much better (+3)}. Ratings were adjusted to reflect processed vs. reference quality and aggregated as Comparison MOS (CMOS).

Six studies were conducted using CCR and ACR-HR with both synthetic and bitrate-ladder datasets on the Prolific platform. Unreliable ratings were removed following ~\cite{naderi_crowdsourcing_2024}, and data were aggregated per degradation condition or bitrate–resolution pair. Average valid votes per condition are reported in Table~\ref{table:n_votes}. While the average standard deviations of both methods are comparable, normalization by the practical scale range shows lower deviation for ACR-HR.

\section{Results}
\label{sec:results}
\subsection{Synthetic impairments}
The correlation coefficients between the opinion scores obtained from different subjective test methods are reported in Table~\ref{table:corr_syn}. In general, higher correlations are observed at the condition level than at the clip level, due to the larger number of votes aggregated together, which reduces noise in the ratings. ACR and ACR-HR exhibit a high degree of correlation, as both rely on the same set of ratings but differ in the calculation method\footnote{Similar correlations were observed with opinion scores obtained from a separate ACR test.}. Correlations are also reported for repeated CCR tests (r1 and r2), demonstrating excellent test–retest reliability without changing the test method. Although the correlations of CCR with both ACR and ACR-HR are comparable for good-quality source videos, ACR-HR shows a significantly stronger correlation with CCR at the clip level than ACR. This finding highlights the necessity of using double-stimulus test methods like CCR or hidden references when evaluating user-generated content. For the subsequent evaluation, only the ACR-HR and CCR test methods are considered. 


Figure~\ref{fig:rd_plots} shows distortion–quality plots for selected degradations. In general, increasing degradation intensity consistently reduced perceived quality. ACR-HR yielded limited scale usage and lower sensitivity to quality drops, particularly for fair-quality source videos (Fig.~\ref{fig:scale}, \ref{fig:scale_dnn}, \ref{fig:scale_dnn_qp}). CCR yielded higher sensitivity and also reflected quality improvements beyond the reference, as observed with super-resolution upscaling (Fig.~\ref{fig:scale_dnn}). Degradations were more pronounced and measurable for good-quality content than for fair-quality content.

\begin{figure}[!t]
    \centering
    
    \subfloat[]{\includegraphics[width=0.87\columnwidth]{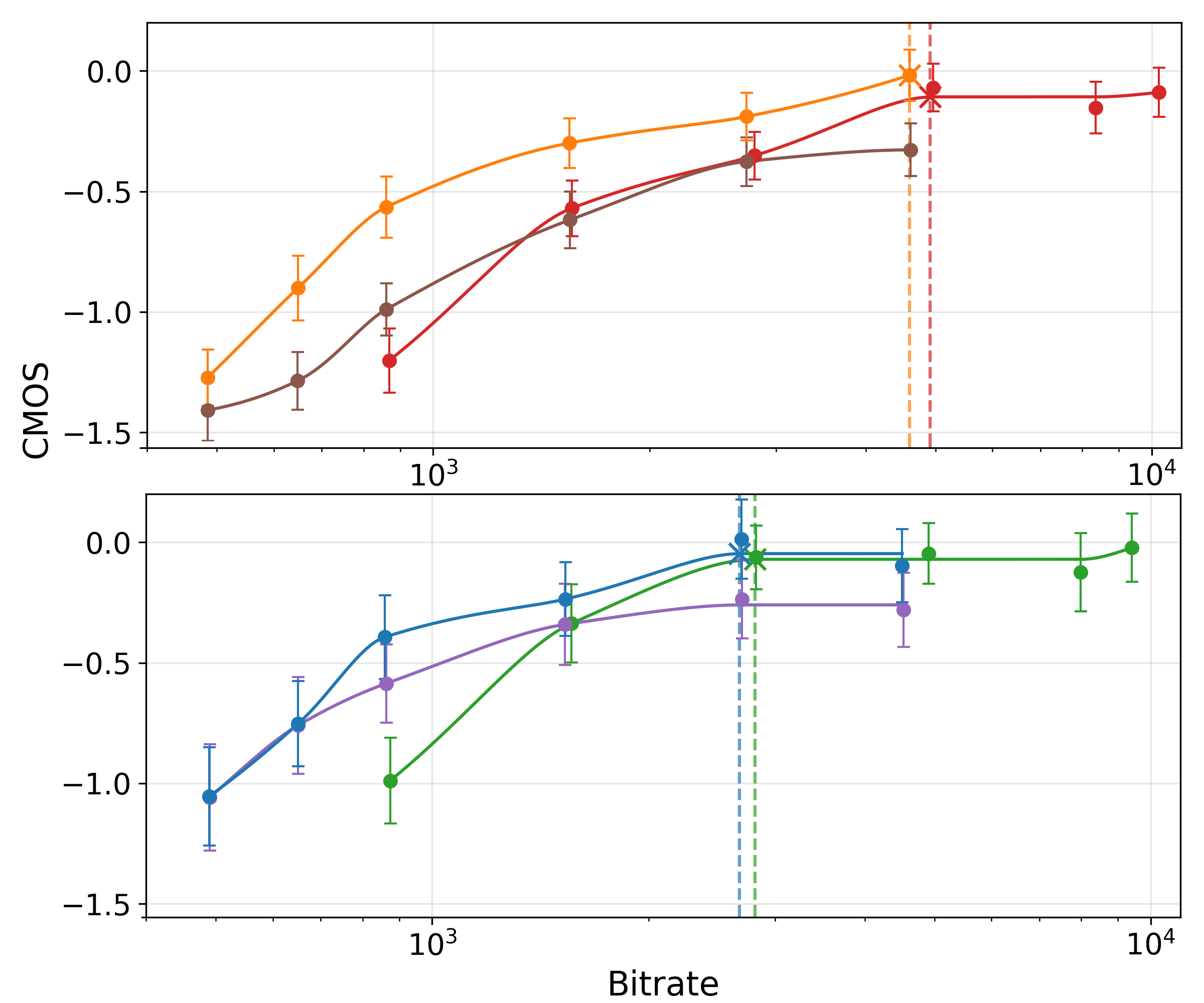}
    \label{fig:bl_cmos}}
    \hfill   
    \subfloat[]{\includegraphics[width=0.85\columnwidth]{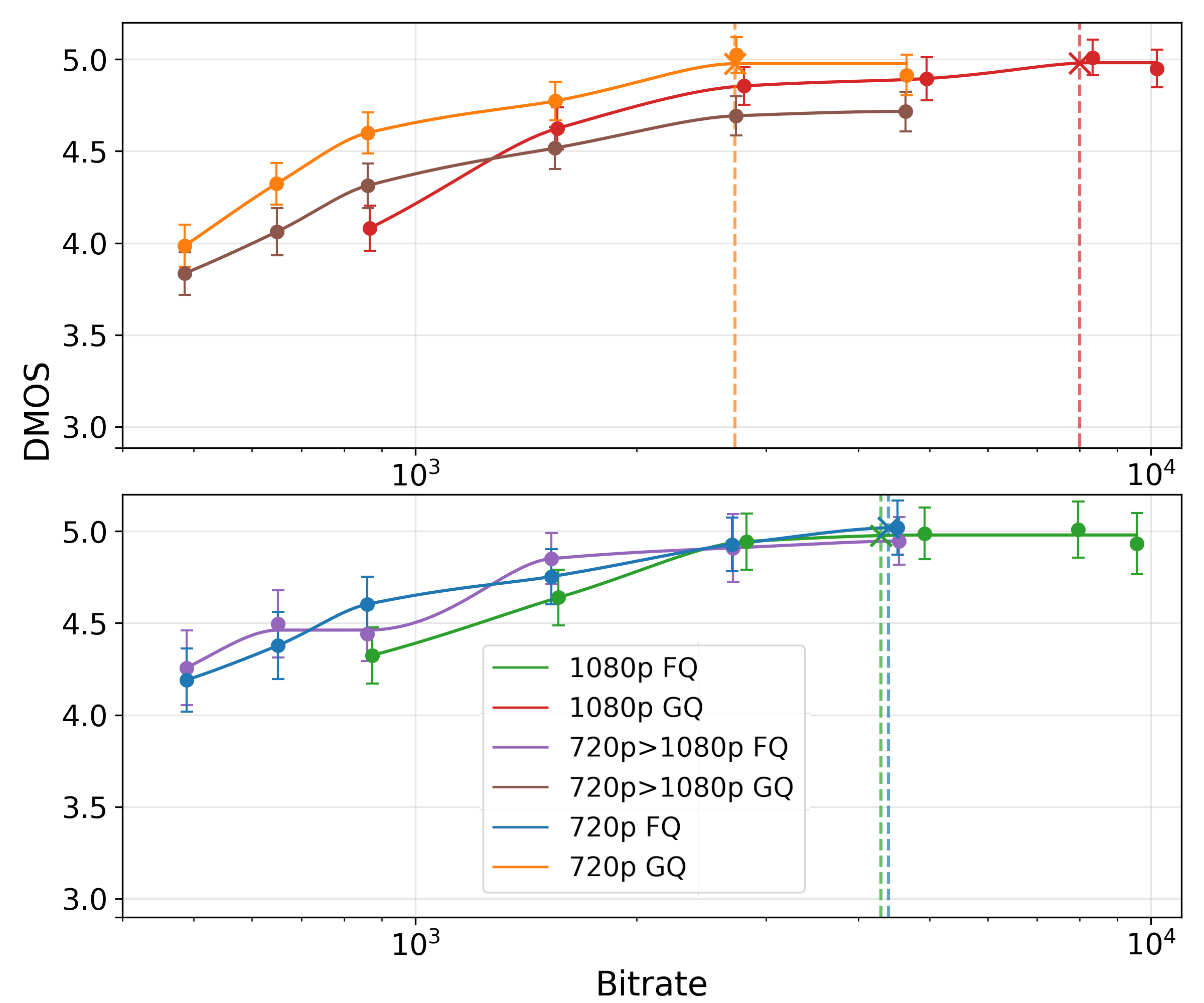}
    \label{fig:bl_dmos}}       
    \hfill   
    \caption{Bitrate ladder for H.264 at two resolutions. Quality was measured with CCR (a) and ACR-HR (b) using Good-Quality (GQ) and Fair-Quality (FQ) sources. Vertical dotted lines indicate quality saturation points for 720p and 1080p.}
    \label{fig:bitrateladder}
    
\end{figure}

\subsection{Bitrate ladder}
Figure~\ref{fig:bitrateladder} presents the rate–distortion plots for three input–target resolutions, measured using the ACR-HR and CCR test methods. Piecewise Cubic Hermite Interpolating Polynomials (PCHIP) were employed to preserve monotonicity when estimating quality saturation points. The bitrates at which quality saturation occurred for each method and source content are reported in Table~\ref{table:saturation}, revealing substantial differences across both test methods and source video qualities.

To further compare the methods, five representative pairs critical for bitrate ladder construction (e.g., 720p videos at 2.7 Mbps against 4.9 Mbps) were evaluated using a Pair Comparison (PC) subjective test. Statistical analysis included the Mann–Whitney U test to assess significant differences between CCR and ACR-HR scores for each pair, and a two-sided binomial test to examine whether PC scores indicated significant differences between pairs. Results show, for good-quality source videos, CCR outcomes were fully consistent with PC, whereas ACR-HR aligned with PC in only 40\% of cases. For fair-quality sources, CCR agreed with PC in 60\% of cases, while ACR-HR achieved agreement in only 20\%.

\begin{table}[h]
\caption{Quality saturation bitrate.}
\label{table:saturation} 
\centering
\resizebox{0.7\columnwidth}{!}{
    \begin{tabular}{l c c c}
    \toprule
    \multirow{2}{*}{\textbf{Input resolution}} & \textbf{Reference} & \multicolumn{2}{c}{\textbf{Bitrate (Kbps)}}  \\    
    & \textbf{quality}&  ACR-HR &  CCR\\
    \midrule        
    \multirow{2}{*}{1080p} & Good & 8002 & 4916 \\
    & Fair & 4296 & 2813 \\
    \multirow{2}{*}{720p} & Good & 2718 & 4603 \\
    & Fair & 4395 & 2685 \\
    
 \bottomrule    
\end{tabular}
}
\vspace{-0.5cm}
\end{table}

\section{Discussion and Conclusion}
\label{sec:discussion}
The ACR-HR test method was found to be approximately twice as fast and cost-effective as the CCR method. It also produced smaller standard deviations when normalized to the practical range of the utilized scale. However, the two methods often led to significantly different conclusions. CCR demonstrated the ability to capture quality enhancements of processed videos beyond the reference, whereas ACR-HR exhibited limited scale range, particularly for fair-quality source videos. Furthermore, the quality of source material influenced outcomes: for blur and scaling artifacts, conclusions differed between good- and fair-quality content, while only minor differences were observed in the case of frame freezing. This suggests that when distortions are sufficiently distinct, source quality has little effect on the conclusions. For encoding videos with constant quantization parameter, no significant differences were found between the two test methods. In the bitrate ladder analysis, in which videos are encoded with constant bitrate, both the choice of test method and the quality of source videos led to differing conclusions. CCR applied to good-quality sources showed full agreement with direct pair-comparison outcomes. 

Overall, the choice of test method should be guided by the study objectives. CCR should be used in cases involving scaling, blur, restoration, or non–good-quality source videos, as it provides greater sensitivity and reliability. By contrast, ACR-HR is suitable for identifying general trends at lower cost when good-quality source videos are available.

\bibliographystyle{IEEEbib}
\bibliography{IC3-AI}

\end{document}